\documentclass[preprint2]{aastex}

\shorttitle{Organic chemistry in dark clouds L1448 and L183}
\shortauthors{Requena-Torres et al.}

\begin{document}


\title{Organic chemistry in the dark clouds L1448 and L183. A Unique grain mantle
composition.}

\author{Requena-Torres, M.A.\altaffilmark{1}\email{requena@damir.iem.csic.es},
Marcelino, N.\altaffilmark{2}\email{nuria@iram.es},
Jim\'enez-Serra, I.\altaffilmark{1}\email{izaskun@damir.iem.csic.es},
Mart\'in-Pintado, J.\altaffilmark{1}\email{martin@damir.iem.csic.es},
Mart\'in, S.\altaffilmark{2}\altaffilmark{3}\email{martin@iram.es}, 
and Mauersberger, R.\altaffilmark{2}\email{mauers@iram.es}}

\altaffiltext{1}{Departamento de Astrof\'isica Molecular e iInfrarroja-
Instituto de Estructura de la Materia-CSIC,c. Serrano 121, E-28006 Madrid, Spain}
\altaffiltext{2}{Instituto de Radioastronom\'ia Milim\'etrica,
Av. Divina Pastora 7, Local 20, E-18012 Granada, Spain}
\altaffiltext{3}{Now at CfA}

\begin{abstract}
We present high sensitivity observations of the complex organic molecules (COMs)
CH$_3$OH, C$_2$H$_5$OH, HCOOCH$_3$, HCOOH and H$_2$CO and of SiO toward the
quiescent dark cloud L183 and the molecular outflow L1448--mm. 
We have not detected C$_2$H$_5$OH, HCOOCH$_3$ and SiO in L183 and in the
quiescent gas of L1448--mm. 
The abundances of CH$_3$OH, 
H$_2$CO and SiO are enhanced by factors 4--20 in the shock precursor component,
and those of CH$_3$OH and SiO by 3 and 4 orders of magnitude in the shocked gas,
without substantial changes ($<$ factor of 2) in the abundances of
C$_2$H$_5$OH and HCOOCH$_3$ relative to that of CH$_3$OH. The large 
enhancements of
SiO and CH$_3$OH can be explained by the shock ejection of an
important fraction of the grain mantle material into gas phase. Our upper limits
to the C$_2$H$_5$OH/CH$_3$OH and HCOOCH$_3$/CH$_3$OH ratios are 
consistent with the rather constant ratios measured in hot cores and
Galactic center clouds. However, the upper limits to the HCOOCH$_3$/CH$_3$OH and
the HCOOH/CH$_3$OH
ratios are at least one order of magnitude smaller than those found in``hot
corinos" surrounding low mass protostars. We speculate that the observed
abundances of COMs in different objects are consistent 
with a sort of ``universal" grain mantle composition
which is locally changed by the processes of low mass star formation.
\end{abstract}

\keywords{astrochemistry --- molecular data --- ISM: molecules ---
techniques: spectroscopic --- ISM: individual(L1448, L183)}

\section{Introduction}
Large abundances of complex organic molecules (COMs) like CH$_3$OH,
C$_2$H$_5$OH, HCOOCH$_3$, HCOOH and H$_2$CO, have been observed in the
interstellar medium (ISM)
mostly associated with grain chemistry \citep{cha95,hor04} in hot
cores associated with massive star formation regions 
\citep{ike01,bis06}, in hot cores associated with low mass star 
formation \citep[the so called ``hot corinos'';][]{bot06}, and
in molecular clouds
affected by shocks in the Galactic center (GC) region \citep{marpin99,req06}.
The comparison between the abundances of COMs in those sources has
shown that the chemistry of COMs in the GC clouds and in hot cores are very
similar suggesting that
evaporation/sputtering of a ``universal'' grain mantle composition are
responsible for the gas phase abundances of complex molecules in these objects.
However, in hot corinos the COMs show different relative abundances
\citep{bot06} than those found in hot cores and GC clouds, suggesting that
either the grain mantle composition in dark clouds is different than in hot
cores and in the GC, or star formation processes locally change the gas
phase abundances of COMs in dark clouds \citep{gar06}.

In dark
clouds without signs of star formation, the abundances of COMs in gas phase
should not be influenced by the evaporation/ejection of complex molecules from
the dust grains, and their abundances should reflect the gas phase chemistry
\citep{irv87,has93a,has93b}. Moreover, observations of solid phase molecules in dark clouds have shown the existence of COMs like HCOOH on the grains mantles 
\citep{kne05}. Dark clouds with very young outflows
should show the abundances of the COMs just ejected from the grain
mantles by shocks.
Therefore dark clouds offer a unique possibility to
study the formation of COMs in the ISM. \\

\citet{fri88} 
have previously observed CH$_3$OH and (CH$_3$)$_2$O in the dark cloud L183, 
deriving upper limits to their relative abundance in agreements with those 
obtained by \citet{req06}.
In this paper, we present high sensitivity
observations of COMs toward the dark clouds L183 and L1448.
L183 is a cold quiescent dark cloud without signatures of
outflow activity and with a rich oxygen chemistry 
dominated by gas phase reactions \citep{pag05}.
L1448 is a dark cloud which harbors an extremely young outflow, L1448--mm. It 
has been proposed that the three velocity components observed in this object 
correspond to the quiescent gas (4.7~km s$^{-1}$), the magnetic precursor of 
C--shocks (5.2~km s$^{-1}$) and the shocked gas 
\citep[broad velocity wings;][]{jim04,jim05}. The enhancement of the SiO and
CH$_3$OH abundances in the shock precursor suggests that these molecules have
been recently ejected from grains with little post ejection processing.
This provides an unique opportunity to establish the
``universality'' of the grain mantle composition.\\  

\section{Observations and results}

The observations of the molecular lines listed in Table \ref{tran} toward L183
and two 
positions in the L1448--mm outflow were carried out with the IRAM 30m radio
telescope at Pico Veleta (Spain).
The data were obtained in two different sessions in 2005. The
half-power beam width of the telescope was $\sim$24$''$, 17$''$ and 12$''$ for
the 3, 2 and 1.3$\,$mm bands.
The receivers, equipped with SIS mixers, were tuned to single sideband with
image rejections of $\ga\,$10$\,$dB. For the observations of L1448--mm we used 
the frequency and wobbler-switched modes with frequency and
position throws of 7.2 MHz and 240$''$, respectively.
A spectral resolution of $\sim$40$\,$kHz corresponding 
to velocity resolutions of $\sim$0.14, 0.08, and 0.05$\,$km
s$^{-1}$ at the observed frequencies was achieved 
using autocorrelators. For L183 we used the frequency switched
mode 
and the same setup as for L1448--mm for
the 2 and 3$\,$mm lines but with a spectral resolution of 80 kHz at
1.3$\,$mm (velocity resolution of 0.09$\,$km s$^{-1}$).
The typical system temperatures were $\sim$100--650$\,$K.
Spectra were calibrated using the standard dual load system.
The line intensities are given in units of $T_{\rm A}^*$.\\

\begin{figure}
\includegraphics[angle=0,width=6cm]{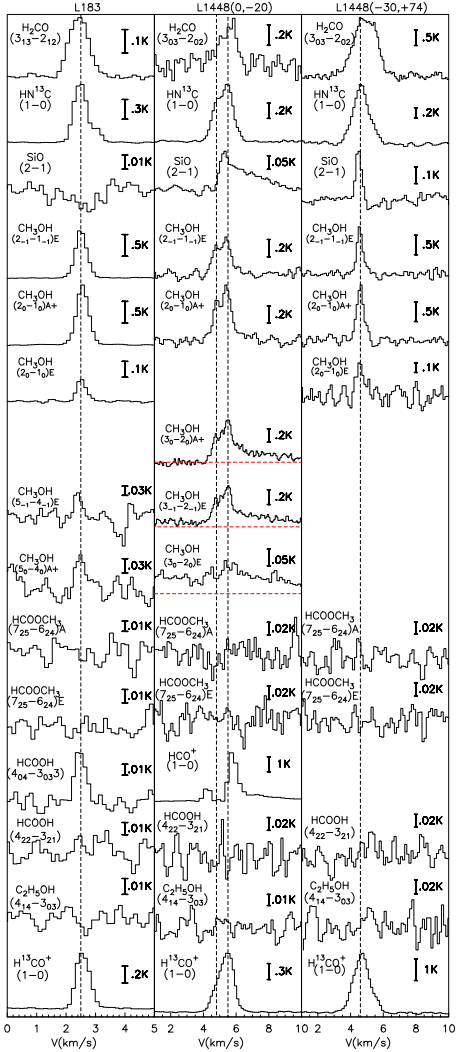}
\caption{Line profiles for the different lines observed toward L183 and L1448,
the dashed vertical lines show the different velocity components. L183
(left panel) shows one-peaked profiles centered at 2.5$\,$km s$^{-1}$.
The line profiles from the L1448~(0,$-$20) (central panel)
and the L1448~($-$30,+74) position (right panel) show velocity components
corresponding to the quiescent gas at 4.8$\,$km s$^{-1}$ and 4.6$\,$km
s$^{-1}$, respectively. The L1448~(0,$-$20)
position also shows the precursor at 5.5$\,$km s$^{-1}$ and, at higher
velocities, the emission associated to the shocked gas for the CH$_3$OH
(3$_0$$\rightarrow$2$_0$)A+ and (3$_{-1}$$\rightarrow$2$_{-1}$)E transitions
at 2$\,$mm.\label{spec}}
\end{figure}

The observed spectra are shown in Figure \ref{spec}. In the left panel we
show the line profiles observed toward L183 and, in the central and right
panels, the line profiles observed toward L1448~(0,$-$20) and ($-$30,+74)
respectively. 
The vertical dashed lines show the quiescent gas in L183 at 2.5$\,$km s$^{-1}$
and the quiescent and the shock precursor components in L1448~(0,$-$20) at 4.8
and 5.2-5.5$\,$km s$^{-1}$. 
For the ($-$30,+74) position in the L1448--mm outflow the the emission is centered at
4.6$\,$km s$^{-1}$ (vertical dashed line in Figure \ref{spec}) and the shock precursor 
as traced by the narrow SiO emission is overlapped with the quiescent gas.
Toward L1448 (0,$-$20), we have
also detected high velocity redshifted wings in CH$_3$OH (2mm lines)
and SiO which are associated with the shocked gas.
The high velocity wings for the  1$\,$mm CH$_3$OH transitions are very
uncertain due to poor baselines in the frequency switching observations.

As expected for a shock tracer,  
the SiO emission is only detected in the shock precursor and
the shocked gas of L1448--mm but not in the quiescent gas of 
L183 and L1448--mm.
CH$_3$OH emission has been detected in all sources and in all velocity
components. H$_2$CO has been detected in all
sources but toward L1448--mm their line profiles are different than the other
molecules. Toward the L1448~(0,$-$20) position, the H$_2$CO emission arises from the
quiescent and shock precursor components like for H$_2$S \citep{jim05}.
Toward the position L1448~($-$30,+74) we also find a broad H$_2$CO profile
indicating the presence of shocked gas. HCOOH has been only
detected in L183 and the most complex molecules like C$_2$H$_5$OH
and HCOOCH$_3$, are not detected toward any of the sources.

The estimated column densities for all
molecules in the different components using the Local Thermodynamic
Equilibrium (LTE) approximation
and the excitation temperatures ($T_{\rm ex}$)
derived from the population diagrams of CH$_3$OH are given in Table \ref{sou3}.
We obtained a
$T_{\rm ex}$ between 5--10$\,$K as expected for dark clouds with
densities of few 10$^{5}$cm$^{-3}$ and kinetic temperatures of 10--20$\,$K
\citep{dic00,cur99}.
We estimated the H$_2$ column densities from the HCO$^+$,
H$^{13}$CO$^+$ and HN$^{13}$C column densities by assuming a
C$^{12}$/C$^{13}$ ratio of 90 and a HCO$^+$/H$_2$ and HNC/H$_2$
abundances of 1$\times$10$^{-8}$ for L1448 \citep{irv87} and a
C$^{12}$/C$^{13}$ ratio of 64 and a HCO$^+$/H$_2$ and HNC/H$_2$
abundances of 8$\times$10$^{-9}$ for L183 \citep{dic00}.  
The H$_2$ column densities and the relative molecular abundances of SiO and
CH$_3$OH are shown in Table \ref{sou3}. This table also shows the abundances of
C$_2$H$_5$OH, HCOOCH$_3$, HCOOH and H$_2$CO relative to that of CH$_3$OH.
              
As expected for the quiescent gas in L1448~(0,$-$20)-q and in L183-q, the SiO
abundance is very low, of $<$2$\times$10$^{-12}$, suggesting that SiO
is depleted onto the grains and on the grain mantles \citep{ziu89,marpin92}.
This is in contrast with the abundance of CH$_3$OH and H$_2$CO in the
quiescent gas which varies by
more than one order and two orders of magnitude, respectively between L1448--mm
and L183. For the other COMs we derive similar upper limits to their
abundances in the quiescent gas in both sources.\\
 
As reported by \citet{jim05}, we find a moderate increase (by a factor of 3) of
the CH$_3$OH abundance and a large increase by a factor of $>$15 in the SiO abundance
in the shock precursor (L1448~(0,$-$20)-p) with respect to that in the quiescent gas.
We also find a large increase by more that one order of magnitude in the H$_2$CO
abundance in the shock precursor in both positions of L1448--mm, as expected if this
molecule, like SiO and CH$_3$OH, was ejected from the grain mantles.
The upper limits to the abundances for
the other COMs are similar to those in the quiescent gas suggesting that the
abundances of these molecules have not been strongly affected by the ejection
from grain mantles.

The abundances of SiO and CH$_3$OH in the shocked gas [see source 
L1448~(0,$-$20)-s in Table \ref{sou3}]
have been enhanced by nearly 4 orders of magnitude with
respect to those of the quiescent gas due to the
sputtering of these molecules from the grain mantles. Surprisingly, the H$_2$CO
to CH$_3$OH abundance ratio decreases by more than one order of magnitude in
the shocked gas as
compared with that in the shock precursor component. Our data suggest that the
abundance of H$_2$CO in the grain mantles is smaller than that of CH$_3$OH.
However, our upper limits to the abundance ratios of C$_2$H$_5$OH, HCOOCH$_3$,
and HCOOH to that of CH$_3$OH in the
shocked gas are small compared with those measured in ``hot
corinos", but consistent with those measured in hot cores and the GC clouds.

\section{Discussion}

The abundances of SiO and of the COMs that we have derived in dark clouds with
different star formation activity show a clear trend.
The quiescent gas in L183 and L1448--mm show a very low 
SiO abundance ($<$10$^{-12}$) as
expected if the molecular complexity in gas phase has not been significantly
affected by the ejection of molecules from grain mantles \citep{marpin92}.
Gas phase models should then explain the observed molecular abundance of
COMs in the quiescent gas. Chemical models for the formation of COMs in dark
clouds \citep{has92,has93a,has93b} predict abundances that are consistent with 

our upper limits for the
COMs in the quiescent gas of L1448--mm, 
except for  H$_2$CO, which shows lower abundances than expected from
chemical models. Low abundances of H$_2$CO ($\sim$10$^{-10}$) are also found
in the envelope of the central position in L1448--mm, possibly related with the
CO depletion \citep[see ][]{mar05}. In the case of
L183, we find discrepancies by up to one order of
magnitude between the predictions and the measured CH$_3$OH, 
HCOOH and H$_2$CO abundances in the quiescent gas.
However, the models of gas-phase chemistry in dark clouds need to be 
revisited since it has been shown that the formation of CH$_3$OH in gas 
phase is much less efficient that previously assumed (see e.g. \citet{gep05}). 
On the other hand, it is very likely that some of the COMs in dark clouds 
are also formed on grains \citep{kne05}.

The SiO abundance in the shock precursor component toward the L1448~(0,$-$20)  position
(5.2$\,$km s$^{-1}$) increases by more than one order of magnitude due to the ejection of
SiO or Si from the grain mantles produced by the sudden acceleration  of ions
relative to neutrals as the C--shocks propagates through the quiescent molecular
cloud \citep{jim04,jim05}. The H$_2$CO abundance in this component is also enhanced by
more than a factor 10. A smaller enhancement by only a factor of $\sim$3 is
observed in CH$_3$OH. The upper limits to the abundance of HCOOH,
C$_2$H$_5$OH and HCOOCH$_3$ in the shock precursor indicate that these molecules are
not ejected from the grain mantles in the first stages of the shock
interaction. A moderate increase
of the CH$_3$OH abundance is also observed for the mixture of quiescent and
precursor gas in L1448~($-$30,+74) together with a large enhancement of the SiO and
H$_2$CO abundances and no enhancement for the other COMs.  

The trend observed in the abundance of some molecules in the shock precursor is
even more
dramatic in the post shocked gas in the L1448--mm outflow.
While the SiO and the CH$_3$OH abundances
increase by 4 and 3 orders of magnitude respectively in the shocked gas, the
abundance ratio of the other COMs (including H$_2$CO) relative to that of  CH$_3$OH
remains similar to those observed in the quiescent gas.

For comparison with our results, Table \ref{sou3} also shows the ratio between the
abundance of COMs to that of CH$_3$OH for all kinds of objects where these
molecules have been detected: hot cores, ``hot corinos'' and molecular
clouds in the GC. The abundance ratios C$_2$H$_5$OH/CH$_3$OH and
HCOOCH$_3$/CH$_3$OH in the GC and in hot cores suggest a similar
grain mantle composition in these two type of objects. In fact, \citet{cha05} have 
proposed paths to form large molecules on the grain mantles which will lead to a 
sort of universal grain mantle composition.

Our upper limits to the relative abundance of
C$_2$H$_5$OH and HCOOCH$_3$ to that of CH$_3$OH in the quiescent and the shocked gas are
consistent with those measured in hot cores and the CG clouds and
therefore, compatible with similar grain mantle composition in all objects.
However, the large COMs abundance ratios measured in hot corinos exceed by more than
one order of magnitude those of the shocked gas in L1448--mm.
Unless  a dramatic change 
in the grain mantle composition between different clouds exists, this suggests that
the large  enhancement of the COMs in gas phase is restricted to the
surrounding material affected by the process of  low mass star formation.

\citet{gar06} have recently proposed that the protostellar 
switch-on phase could
lead to a very effective grain-surface and gas formation of COMs due
to the warm up
of the circumstellar gas and dust.
Although the models are not able to reproduce the large HCOOCH$_3$/CH$_3$OH
ratio found in hot corinos they indicate local changes associated with star
formation.
The large abundance of COMs observed in the GC clouds cannot
be however explained by this model, since star formation does not play any role
over large scales, and the grain surface reactions are not very
effective due to relatively low temperatures for the 
bulk of grains \citep[$<$15$\,$K,][]{rod04}.\\

In summary, the low abundance ratios of COMs that we find in the two dark
clouds, support the scenario proposed by \citet{mar05} where the abundances are locally
enhanced in the circumstellar material. But the bulk of the quiescent material could 
have the
same grain mantle composition as observed in the GC clouds and in the hot
cores, suggesting a sort of ``universal'' grain mantle composition as proposed by
\citet{req06}. High sensitivity and resolution measurements of the COMs abundances 
towards the exciting source of the L1448--mm outflow will help to clarify if the
high abundances of some of the COMs observed in hot corinos are a
local effect associated with low mass star formation.

\acknowledgments
We wish to thank C. Ceccarelli for her comments and S. Bisschop and her co-authors
for communicating the results of their work prior to publication. 
This work has been supported
by the Spanish Ministerio de Educaci\'on y Ciencia under projects ESP~2004-00665,
AYA~2003-02785, and ``Comunidad de Madrid"
Government under PRICIT project S-0505/ESP-0237 (ASTROCAM).

\clearpage

\begin{deluxetable}{lcccccccc}
\tabletypesize{\tiny}
\tablewidth{0pc}
\tablecolumns{3}
\tablecaption{Observed transitions\label{tran}}
\tablehead{\colhead{Molecules}&\colhead{Transition}
&\colhead{frequency}&\colhead{E$_u$}\\
\colhead{}&\colhead{}&\colhead{MHz}&\colhead{K}}
\startdata
HCO$^+$.....        &$1$$\rightarrow$$0$        &89187.41&4.29\\
H$^{13}$CO$^+$....&$1$$\rightarrow$$0$        &86754.33&4.17\\
HN$^{13}$C....        &$1$$\rightarrow$$0$        &87090.85&4.2\\
C$_2$H$_5$OH.... &$4_{14}$$\rightarrow$$3_{03}$        &90117.61&9.36\\
SiO........        &$2$$\rightarrow$$1$                &86846.96&6.26\\
HCOOCH$_3$....        &$7_{25}$$\rightarrow$$6_{24}$E        &90145.69&19.69\\
                &$7_{25}$$\rightarrow$$6_{24}$A        &90156.48&19.67\\
HCOOH......        &$4_{22}$$\rightarrow$$3_{21}$        &90164.63&23.53\\
                     &$4_{04}$$\rightarrow$$3_{03}$        &89579.17&13.57\\
H$_2$CO.......        &$4_{22}$$\rightarrow$$3_{21}$         &211211.47&32.07\\
                 &$3_{03}$$\rightarrow$$2_{02}$         &218222.19&20.97\\
CH$_3$OH......         &$3_{0}$$\rightarrow$$2_{0}$A+         &145103.23&13.94\\
                 &$3_{-1}$$\rightarrow$$2_{-1}$E        &145097.47&19.52\\
                 &$3_{0}$$\rightarrow$$2_{0}$E        &145093.75&27.06\\
                 &$2_{0}$$\rightarrow$$1_{0}$A+         &96741.42&6.97\\
                 &$2_{-1}$$\rightarrow$$1_{-1}$E        &96739.39&12.55\\
                 &$2_{0}$$\rightarrow$$1_{0}$E         &96744.58&20.10\\
                 &$5_{0}$$\rightarrow$$4_{0}$E         &241700.22&47.95\\
                 &$5_{-1}$$\rightarrow$$4_{-1}$E        &241767.22&40.41\\
                 &$5_{0}$$\rightarrow$$4_{0}$A+         &241791.43&34.83\\
\enddata
\tablecomments{Quantum numbers and frequencies of the observed transitions
from the Jet Propulsion Laboratory molecular catalog \citep{pic98}.
We observed L183 toward the position
$\alpha$[J2000]=15$^{\rm h}$54$^{\rm m}$08$^{\rm s}$.6,
\mbox{$\delta$[J2000]=-02$^o$52$'$10$''$00} and two different
offsets, (0,$-$20) and ($-$30,+74), in the L1448 cloud  relative to L1448--mm
position ($\alpha$[J2000]=03$^{\rm h}$25$^{\rm m}$38$^{\rm s}$,
$\delta$[J2000]=30$^o$33$'$05$''$).}
\end{deluxetable}        

\clearpage

\begin{deluxetable}{lcccccccc}
\tabletypesize{\tiny}
\tablewidth{0pc}
\tablecolumns{9}
\tablecaption{Derived abundances\label{sou3}}
\tablehead{\colhead{source$^1$}&\colhead{$T_{\rm ex}$}&\colhead{N(H$_2$)}&
\colhead{$X_{\rm SiO}$}&\colhead{$X_{\rm CH_3OH}$}
&$\frac{\rm C_2H_5OH}{\rm CH_3OH}$&$\frac{\rm HCOOCH_3}{\rm CH_3OH}$
&$\frac{\rm HCOOH}{\rm CH_3OH}$&$\frac{\rm H_2CO}{\rm CH_3OH}$\\
\colhead{}&\colhead{(K)}&\colhead{$\times10^{22}$cm$^{-2}$}&\colhead{$\times10^{-12}$}&
\colhead{$\times10^{-9}$}
&\colhead{$\times$10$^{-2}$}&\colhead{$\times$10$^{-2}$}
&\colhead{$\times$10$^{-2}$}&\colhead{$\times$10$^{-2}$}}
\startdata
L183-q                        &  5.7        &  0.4                        & $\la$4                        & 18.4                  & $\la$0.4        & $\la$3        & 1.3                  &12.8       \\  
L1448~(0,$-$20)-q                &  8.9        &  1.2                        & $\la$2                        & 1.4                          & $\la$4        & $\la$15        & $\la$9          &$\la$0.1  \\
L1448~(0,$-$20)-p                &  9.8        &  1.2                        & 29.1                                & 3.9                          & $\la$3        & $\la$9        & $\la$5          &2.2       \\
L1448~($-$30,+74)-q+p        &  9.3        &  3.7                        & 12.5                                & 2.2                          & $\la$1        & $\la$5        & $\la$3         &15.0     \\
L1448~(0,$-$20)-s                &&&&&&&&\\
6--8$\,$km$\,$s$^{-1}$        &15        &2.1$\times$10$^{-3}$        &2.3$\times$10$^{4}$                &1.5$\times$10$^{3}$        &$\la$7                &$\la$17              &$\la$9               &$\la$0.2\\
8--10$\,$km$\,$s$^{-1}$        &15        &7.4$\times$10$^{-4}$        &1.4$\times$10$^{4}$                &2.8$\times$10$^{3}$        &$\la$11        &$\la$25              &$\la$13              &$\la$0.3\\
10--12$\,$km$\,$s$^{-1}$&15        &4.8$\times$10$^{-4}$        &$\la$1.2$\times$10$^{4}$        &2.7$\times$10$^{3}$        &$\la$17        &$\la$40              &$\la$21              &$\la$0.4\\
12--14$\,$km$\,$s$^{-1}$&15        &3.2$\times$10$^{-4}$        &$\la$1.8$\times$10$^{4}$        &2.6$\times$10$^{3}$        &$\la$27        &$\la$63              &$\la$33              &$\la$0.7\\
14--16$\,$km$\,$s$^{-1}$&15        &2.9$\times$10$^{-4}$        &$\la$3.2$\times$10$^{4}$        &3.4$\times$10$^{3}$        &$\la$45        &$\la$76              &$\la$35              &$\la$0.9\\
\hline
GC sources$^2$                &        &                        &                                &                        & $\sim$3.7        &$\sim$3.9             &$\sim$0.8             & $\sim$1.1     \\  
Hot cores$^3$                &        &                        &                                &                        & $\sim$2.5        &$\sim$6.9             &$\sim$0.02$^*$        & $\sim$16.0        \\  
Hot corinos$^4$                &        &                        &                                &                        &                &$\sim$160        &$\sim$60        & $\sim$190      
\enddata
\tablecomments{H$_2$ column density, temperature and the relative abundances
of the different COMs to that of H$_2$ and of CH$_3$OH. $^1$ q for
quiescent, p for precursor, and s for shocked gas. $^2$ \citet{req06}. $^3$
\citet{bis06}. $^4$ \citet{bot06}. $^*$ Assumed by the authors to be related
with a cooler gas than the other complex molecules. }
\end{deluxetable}     
  
\clearpage

\clearpage

\begin{thebibliography}{}
\bibitem[Bachiller \& P\'erez-Guti\'errez(1997)]{bac97} Bachiller, R. \&
P\'erez-Guti\'errez, M. 1997, \apj, 487, L93
\bibitem[Bisschop et al.(2006)]{bis06} Bisschop, S. E., J\o{}rgensen,
J. K., de Wachter, E., \& van Dishoeck, E. F. 2006, \aap, submitted
\bibitem[Bottinelli et al.(2006)]{bot06} Bottinelli, S., Ceccarelli, C.,
Williams, J. P., \& Lefloch, B. 2006, \aap, accepted. astro-ph/0611480.
\bibitem[Boudin et al.(1998)]{bou98} Boudin, N., Schutte, W. A., \& Greenberg, J.
M. 1998, A\&A, 331, 749
\bibitem[Charnley et al.(1995)]{cha95} Charnley, S. B., Kress, M. E., Tielens,
A. G. G. M., \& Millar, T. J. 1995, \apj, 448, 232
\bibitem[Charnley \& Rodgers(2005)]{cha05}Charnley, S. B. \& Rodgers, S. D. 
2005, IAU 231 Symposium book ``Astrochemistry: Recent Successes and Future 
Challenges" Ed. Lis, D. C., Blake, G. A., \& Herbst, E.
\bibitem[Curiel et al.(1999)]{cur99}Curiel, S., Torrelles, J. M., Rodr\'iguez, L. F.,
G\'omez, J. F., \& Anglada, G. 1999, \apj, 527, 310
\bibitem[Dickens et al.(2000)]{dic00} Dickens, J. E., Irvine, W. M., Snell,
 R. L., Bergin, E. A., Schloerb, F. P., Pratap, P., \& Miralles, M. P.
2000, \apj, 542, 870
\bibitem[Friberg et al.(1988)]{fri88} Friberg, P., Madden, S. C., Hjalmarson, \AA.,
\& Irvine, W. M. 1988, \aap, 195, 281
\bibitem[Garrod \& Herbst(2006)]{gar06} Garrod, R. T. \& Herbst, E. 2006,  \aap,
457, 927 
\bibitem[Geppert et al.(2005)]{gep05} Geppert, W. D., Hellberg, F., \"Osterdahl, 
F. et al. 
2005, IAU 231 Symposium book ``Astrochemistry: Recent Successes and Future Challenges"
 Ed. Lis, D. C., Blake, G. A., \& Herbst, E.
\bibitem[Grim et al.(1991)]{gri91} Grim, R. J. A., Bass, F., Geballe, T. R.,
Greenberg, J. M., \& Schutte, W. 1991, A\&A, 243, 473
\bibitem[Hasegawa et al.(1992)]{has92} Hasegawa, T. I., Herbst, E.,
\& Leung, C. M. 1992, \apjs 82, 167
\bibitem[Hasegawa \& Herbst(1993a)]{has93a} Hasegawa, T. I., \& Herbst, E.
1993a, MNRAS, 261, 83
\bibitem[Hasegawa \& Herbst(1993b)]{has93b} Hasegawa, T. I., \& Herbst, E.
1993b, MNRAS, 263, 589
\bibitem[Horn et al.(2004)]{hor04} Horn, A., M\o{}llendal, H.,
Sekiguchi, O. et al. 
 2004, \apj, 611, 605
\bibitem[Ikeda et al.(2001)]{ike01} Ikeda, M., Ohishi, M., Nummelin, A.,
Dickens, J. E., Bergman, P., Hjalmarson, \AA., \& Irvine, W. M. 2001, \apj,
 560, 792
\bibitem[Irvine et al.(1987)]{irv87} Irvine, W. M.,
Goldsmith, P. F., \& Hjalmarson, \AA. 1987, ``Interstellar
processes". Hollenbach, D. J., Tromson, H. A. (eds.)
Reidel Dordrecht, p. 561
\bibitem[Irvine et al.(1991)]{irv91} Irvine, W. M.,
Ohishi, M., \& Kaifu, N. 1991, ICARUS, 91, 2
\bibitem[Jim\'enez-Serra et al.(2004)]{jim04} Jim\'enez-Serra, I.,
Mart\'in-Pintado, J., Rodr\'iguez-Franco, A. \& Marcelino, N. 2004
\apj, 603, L49
\bibitem[Jim\'enez-Serra et al.(2005)]{jim05} Jim\'enez-Serra, I., Mart\'in-Pintado,
J., Rodr\'iguez-Franco, A., \& Mart\'in, S. 2005, \apj, 627, L121
\bibitem[Knez et al.(2005)]{kne05} Knez, C., Boogert, A.C.A., Pontoppidan,  
K. M. et al. 
2005, \apj, 635, L145 
\bibitem[Marcelino et al.(2006)]{marc06} Marcelino et al. 2006, in preparation
\bibitem[Maret et al.(2005)]{mar05} Maret, S., Ceccarelli, C., Tielens, A. G.
M., Caux, E., Lefloch, B., Faure, A., Castets, A., \& Flower, D. R. 2005, A\&A,
442, 527
\bibitem[Mart\'in-Pintado et al.(1992)]{marpin92} Mart\'in-Pintado, J.,
Bachiller, R., \& Fuente, A. 1992, \aap, 254, 315
\bibitem[Mart\'in-Pintado et al.(1999)]{marpin99} Mart\'in-Pintado, J., Gaume,
R. A., Rodr\'iguez-Fern\'andez, N. J., de Vicente, P., \& Wilson, T. L. 1999,
 \apj, 519, 667
\bibitem[Pagani et al.(2005)]{pag05} Pagani, L., Pardo, J.-R., Apponi, A. J.,
Bacmann, A., \& Cabrit, S. 2005, \aap, 429, 181
\bibitem[Pickett et al.(1998)]{pic98} Pickett, H. M., Poynter, R. L., Cohen, 
E. A., Delitsky, M. L., Pearson, J. C., Muller, H. S. P., 1998,``Submillimeter, 
Millimeter, and Microwave spectral line Catalog", J. Quant. spectrosc. \& Rad. 
Transfer 60, 883
\bibitem[Requena-Torres et al.(2006)]{req06} Requena-Torres, M. A., Mart\'in-Pintado, J.,
Rodr\'iguez-Franco, A., Mart\'in, S., Rodr\'iguez-Fern\'andez, N. J., \& de Vicente, P.
2006, \aap, 455, 971
\bibitem[Rodr\'iguez-Fern\'andez et al.(2004)]{rod04} Rodr\'iguez-Fern\'andez,
N. J., Mart\'in-Pintado, J., Fuente, A., \& Wilson, T. L. 2004, \aap, 427, 217
\bibitem[Tielens(1992)]{tie92} Tielens, , A. G. G. M. 1992, in ``Chemistry and
Spectroscopy of Interestellar Molecules", ed. N. Kaifu (Tokyo: Univ. Tokyo
Press), 237
\bibitem[Watanabe et al.(2003)]{wat03} Watanabe, N., Shirak, T., \& Kouchi, A. 2003, \apj, 588, L121
\bibitem[Ziurys et al.(1989)]{ziu89} Ziurys, L. M., Friberg, P. \& Irvine, W.
M. 1989, \apj, 343, 201
\end{thebibliography}
\end{document}